\documentstyle[sprocl,epsfig]{article}

\def\nostrocostrutto#1\over#2{\mathrel{\mathop{\kern 0pt \rlap 
  {\raise.2ex\hbox{$#1$}}}
  \lower.9ex\hbox{\kern-.190em $#2$}}}
\def\lsim{\nostrocostrutto < \over \sim}   

\newcommand{\be}{\begin{equation}}
\newcommand{\ee}{\end{equation}}
\newcommand{\ba}{\begin{eqnarray}}
\newcommand{\ea}{\end{eqnarray}}

\newcommand{\eref}[1]{(\ref{#1})}      

\begin{document}
\thispagestyle{empty}
\setcounter{page}{0}
\noindent
\rightline{Durham DTP/96/94}
\rightline{MPI-PhT/96-94}
\rightline{October 1996}
\vspace{1.0cm}
\begin{center}
{\Large \bf Soft particle production and QCD coherence} 
\end{center}
\vspace{0.5cm} 
\begin{center}
VALERY A. KHOZE~$^{a, b,}~$\footnote{e-mail: v.a.khoze@durham.ac.uk} \ , \ 
SERGIO LUPIA~$^{c,}~$\footnote{e-mail: lupia@mppmu.mpg.de} \ , \ 
WOLFGANG OCHS~$^{c,}~$\footnote{e-mail: wwo@mppmu.mpg.de} 
\end{center}

\medskip 

\begin{center}
$^a$ \ {\it Department of Physics\\
University of Durham, Durham DH1 3LE, UK}\\ 
\mbox{ }\\
$^b$ \ {\it Institute for Nuclear Physics\\
St. Petersburg, Gatchina, 188350, Russia}\\
\mbox{ }\\
$^c$ \ {\it Max-Planck-Institut f\"ur Physik \\
(Werner-Heisenberg-Institut) \\
F\"ohringer Ring 6, D-80805 Munich, Germany} 
\end{center}
\vspace{1.0cm}

\begin{abstract}
We discuss the behaviour of the energy spectrum of particles in jets 
near the limit of small momenta of a few hundred MeV. In QCD parton 
cascades the soft gluons are coherently emitted from all faster partons in
the jet and their production rate is predicted to scale. The observed
charged and identified particle spectra follow this behaviour surprisingly
well supporting of the hypothesis of the Local Parton Hadron Duality (LPHD) 
for this extreme limit. Further tests of this perturbative approach are
 discussed.
\end{abstract}

\vfill 
\noindent 
to appear in the Proc. of the 7th International Workshop ``Correlations and
Fluctuation'' (Nijmegen, the Netherlands, June 30-July 6, 1996)
\vfill 

\newpage

\title{SOFT PARTICLE PRODUCTION AND QCD COHERENCE} 

\author{VALERY A. KHOZE}
\address{Department of Physics,
University of Durham, Durham DH1 3LE, UK\\
and\\
 Institute for Nuclear Physics,
St. Petersburg, Gatchina, 188350, Russia}
\author{SERGIO LUPIA \ and \
WOLFGANG OCHS}
\address{Max-Planck-Institut f\"ur Physik (Werner-Heisenberg-Institut) \\
F\"ohringer Ring 6, D-80805 Munich, Germany} 

\maketitle
\abstracts{
We discuss the behaviour of the energy spectrum of particles in jets 
near the limit of small momenta of a few hundred MeV. In QCD parton 
cascades the soft gluons are coherently emitted from all faster partons in
the jet and their production rate is predicted to scale. The observed
charged and identified particle spectra follow this behaviour surprisingly
well supporting of the hypothesis of the Local Parton Hadron Duality (LPHD) 
for this extreme limit. Further tests of this perturbative approach are
 discussed.} 

\section{Introduction}

The study of multiparticle production in hard collision processes can yield
valuable information about the characteristic features of the partonic
branching processes in QCD and the transition from the coloured partons to
the colourless hadrons. The parton branching process is driven by the gluon
bremsstrahlung which becomes singular in the limit of collinear or soft
emission. Analytical calculations in the Double Logarithmic Approximation (DLA)
take into account the contributions from these singularities. An improved
accuracy is obtained in the Modified Leading Logarithmic Approximation
(MLLA) which includes corrections of relative order $\sqrt{\alpha_s}$
(for a review, see \cite{dkmt2}).

There is no commonly accepted theory of hadronization, but experimental data
can be used to improve our knowledge about the partonic interactions
in the limit of small momentum transfer. Here we consider the hypothesis of
the LPHD~\cite{adkt1} which assumes that
sufficiently inclusive observables at the hadronic level are directly given
by the corresponding quantities computed for the parton cascade. 
A striking success of this approach is the prediction of the energy
distribution of particles with an approximately Gaussian shape in the 
variable $\xi = \log E_{jet}/E$ for particles
with energy $E$ in a jet of energy $E_{jet}$ (the so-called ``hump-backed
plateau'')\cite{adkt1,dfk1,bcmm}.
 
Here we focus on the soft end of the spectrum ($E\lsim 1$ GeV). 
We recall that the gluons 
of  long wave length are emitted coherently by
the total colour current which is independent of the internal structure of the
jet and is conserved when the partons split. The soft radiation then depends
essentially on the total colour charge of the initial partons. 
Due to the coherence of the gluon radiation it is not the softest partons but
those with intermediate energies ($E \sim E_{jet}^{0.3-0.4}$) which 
multiply most effectively in QCD cascades. 
Applying the LPHD hypothesis
one expects that the hadron spectrum at low momentum $p$ should be 
nearly independent of
the jet energy $E_{jet}$\cite{adkt1,vakcar,lo}.
Quantitatively, we consider the invariant density $E dn/d^3p\equiv dn/dy
d^2p_T$ in the limit of vanishing rapidity $y$ and transverse momentum
$p_T$, or, equivalently, for vanishing momentum $p$, i.e.
\be
I_0 = \lim_{y \to 0, p_T \to 0} E \frac{dn}{d^3p} = 
\frac{1}{2} \lim_{p \to 0} E \frac{dn}{d^3p}
\label{izero}
\ee
where the factor 1/2 takes into account that both hemisphere are added in
the
limit $p \to$ 0. 
If the dual description of hadronic and partonic final states is really
adequate down to very small momenta, the finite, energy independent limit of
the invariant hadronic density $I_0$ is expected 
from the colour coherence argument.
This expectation is supported by the experimental data
for both charged and  identified hadrons
in the full energy range explored so far in $e^+e^-$ annihilation.
Furthermore we discuss how the QCD picture explaining this result can be
 further tested. For more details we refer to the paper \cite{klo2}.

\section{Particle production in the soft limit}

We consider first the analytical predictions for the energy spectrum of
partons near the soft limit. The asymptotic behaviour of the energy spectrum
is obtained from the DLA in which energy
conservation is neglected and only the leading singularities in the parton
splitting functions are kept. 
The evolution equation of the single parton inclusive energy 
distribution originating from a primary parton $A$ is given by\cite{dfk1}: 
\be
D_A^p(\xi,Y) = \delta_A^p\delta(\xi) + \int_0^{\xi} d\xi' \int_0^{Y-\xi} 
dy' \frac{C_A}{N_C} \gamma_0^2(y') D_g^p(\xi',y') \; . 
\label{evoleq}
\ee
Here  we have used the logarithmic variables 
 $\xi = \log (1/x) = \log (Q/E)$ and $Y = \log (Q/Q_0)$ with
$E$ the particle energy and $Q$ the jet virtuality ($Q=P\Theta$ for
a jet of primary momentum $P$ and opening angle $\Theta$); 
 $C_A$ is the respective colour factor, $N_C$ for $A=g$ and $C_F$ for $A=q$;
 $\gamma_0$ denotes the anomalous dimension of multiplicity and 
is related to the 
QCD running coupling by $\gamma_0^2 = 4 N_C \alpha_s / 2 \pi$
or $\gamma_0^2 = \beta^2/\log(p_{\perp}/\Lambda)$ with $\beta^2=4 N_C /b$,
$b \equiv (11 N_c - 2 n_f)/3$;  
$\Lambda$ is the QCD-scale and  $N_C$ and $n_f$ are 
the number of colours and of flavours respectively. 
The shower evolution is cut off by 
$Q_0$, such that the transverse momentum $p_{\perp} \ge Q_0$. 


In case of fixed $\alpha_s$  the exact DLA solution is known\cite{dfk1}
\begin{eqnarray}
D_A^g(\xi,Y,\gamma_0) & = & \delta_A^g \delta(\xi)+
\frac{C_A}{N_C} \gamma_0 \sqrt{\frac{Y-\xi}{\xi}} 
I_1 \biggl(2 \gamma_0 \sqrt{\xi (Y-\xi)} \biggr)
\label{fixed:solution}\\
  & \approx &  \delta_A^g \delta(\xi)+
\frac{C_A}{N_C} \gamma_0^2 (Y-\xi)
    \biggl(1+\frac{1}{2} \gamma_0^2 \xi (Y-\xi) + \ldots\biggl)
\nonumber
\end{eqnarray}
where $I_1$ is a modified Bessel
function of first kind. This distribution in $\xi$ vanishes for $\xi\to 0$
and $\xi\to Y$ as required by eq. (\ref{evoleq}). The exact MLLA correction
is obtained \cite{lo} (see also \cite{dt}) by multiplying the
DLA result (\ref{fixed:solution}) with the factor
$\exp(- a\gamma_0^2(Y-\xi)/4N_C)$ where 
 $a=\frac{11}{3}N_C+\frac{2n_f}{3N_C^2}$.

In case of running $\alpha_s$ the first terms of an iterative solution of
the DLA equation (\ref{evoleq}) have been derived; the MLLA correction is
again given by an
exponential damping factor  in analogy to
the fixed $\alpha_s$ case \cite{klo2}.

While the limiting behaviour of the partonic energy spectrum in the soft
region follows from the general principle of colour coherence the detailed
form of the observable hadronic spectrum is predicted uniquely from the 
LPHD hypothesis only for $E \simeq p \gg Q_0$, but not  near the 
kinematical boundary because of the sensitivity of the spectrum 
to the cut-off procedure. In the usual application of the 
MLLA the partons are
treated as massless with energy $E=p\geq p_{\perp} \geq Q_0$, so $\xi\leq Y$.
Experimental hadronic spectra are usually presented as function of momenta
$p$ or $\xi_p=\log (1/x_p)$ which is not limited from above. The same
kinematic limit for partons and hadrons is obtained if the hadronic mass
$m_h$ and the partonic $p_{\perp}$ cut-off $Q_0$ are taken the same.

For the relation between parton and hadron distributions one may require
that the invariant density $E dn/d^3p$ of hadrons approaches a constant limit
for $p\to 0$ as is observed experimentally. For the spectra which
vanish linearly as in (\ref{fixed:solution}) this is achieved by relating 
the hadron and parton spectra as \cite{dfk1,lo}
\be
E_h \frac{dn(\xi_E)}{dp_h} = K_h E_p \frac{dn(\xi_E)}{dp_p}
    \equiv K_h D_A^g(\xi_E,Y)
    \label{phrel}
\ee
with $E_h=\sqrt{p_h^2+Q_0^2}=E_p \geq Q_0$, where $K_h$ is a normalization
parameter referring to a single jet. 
Then, indeed, for hadrons the invariant density
$E dn/d^3p=K_h D_A^g(\xi,Y)/4\pi (E_h^2-Q^2_0)$
approaches the finite limit as in (\ref{izero})
\be
I_0=K_h \frac{C_A\beta^2}{8\pi N_C\lambda Q_0^2}.
\label{limit}
\ee 
In the fixed $\alpha_s$ limit $\beta^2/\lambda$ is replaced by $\gamma_0$
and $I_0\sim 1/Q^2_0$.
With prescription (\ref{phrel}) the moments of the full energy spectrum 
$D(\xi,Y)$ are
well described by the MLLA formulae at $Q_0=270$ MeV in a wide energy
range \cite{lo,klo1}.

The relation (\ref{phrel}) is not unique, however. We found that the
alternative prescription based on phase space arguments, 
$dn/d\xi_p = (p/E)^3 D(\xi,Y)$  (see e.g.~\cite{DKTInt}): 
\be
E_h\frac{dn}{d^3p_h} = K_h \biggl(\frac{1}{4 \pi E^2}\biggr) D_A^g(\xi_E,Y)
\label{phrel1}
\ee
works well for charged pions for all energies $E$ in the LEP region and for 
charged particles at low energies $E$ 
if $D_A^g$ is the MLLA ``limiting spectrum" with
$Q_0=\Lambda=138$ MeV. The low cut-off mass is plausible in this region,
which is dominated by pions. 

\begin{figure}
          \begin{center}
\mbox{\epsfig{file=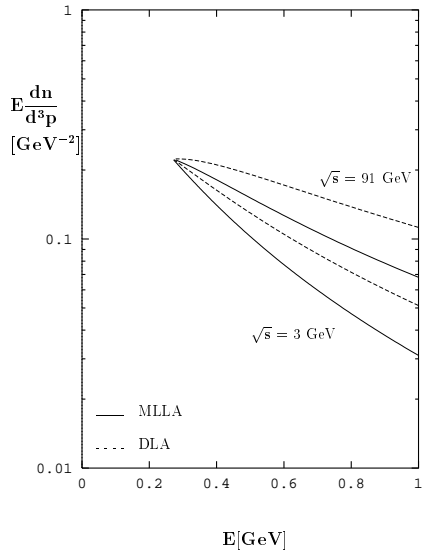,bbllx=4.5cm,bblly=10.cm,bburx=16.5cm,bbury=27.cm,height=7cm}}
       \end{center}
\caption{Invariant density $E dn/d^3p$ 
as a function of the particle energy $E$ for $Q_0$ = 270 MeV. 
Predictions of MLLA and DLA with fixed $\alpha_s$ (with $\gamma_0$ = 0.64) 
at $cms$ energies of $\protect\sqrt{s}$ = 3 GeV (lower two curves) and 91 
GeV (upper two curves)  with $D_g^g$ computed 
using eq.~\protect\eref{phrel}.}
\label{fixed}
\end{figure}

In order to illustrate the above exact 
analytical results we compare in Fig. 1
the computations from DLA and MLLA (in case of fixed
$\alpha_s$) for low particle energies
using the relation (\ref{phrel}) between parton and hadron spectra.
The single particle invariant density
 approaches an
energy independent value in the soft limit $\xi\to Y$ ($E\to Q_0$). This
originates from the soft gluon emission contribution of order $\alpha_s$
(the term of order $\gamma_0^2$ in the expansion (\ref{fixed:solution}))
which is determined
by the total colour charge of the primary partons due to the colour
coherence. In this limit the MLLA converges towards the DLA as the energy
conservation constraints are unimportant and the parton splitting functions
are only probed for very small fractional momenta. The approximate results
with running $\alpha_s$ show the same features \cite{klo2}.

\section{Discussion of experimental results}

\begin{figure}
\vfill \begin{minipage}{.45\linewidth}
          \begin{center}
\mbox{\epsfig{file=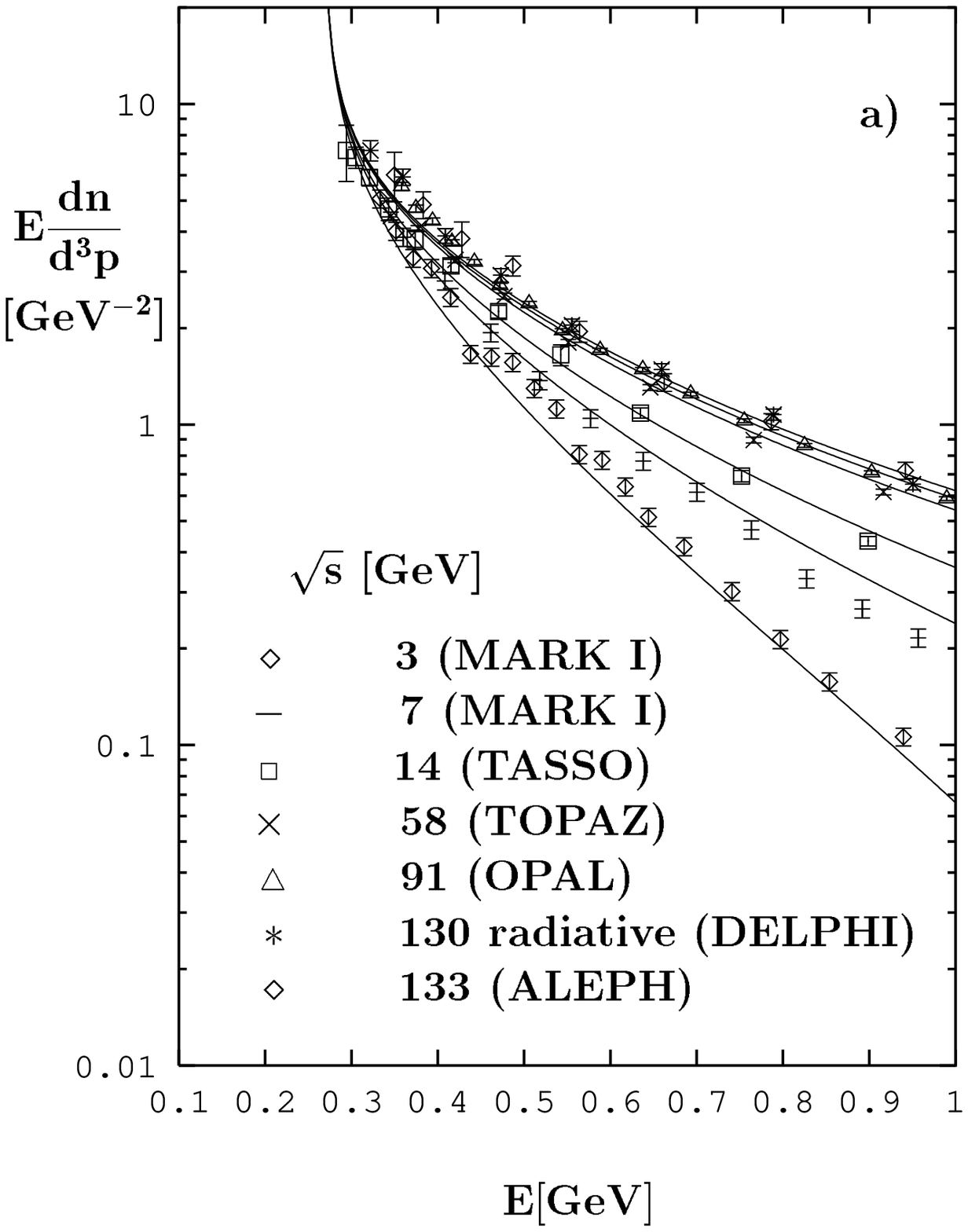,width=.6\linewidth,bbllx=5.5cm,bblly=10.5cm,bburx=13.5cm,bbury=26.5cm}}
          \end{center}
      \end{minipage}\hfill
      \begin{minipage}{.45\linewidth}
          \begin{center}
\mbox{\epsfig{file=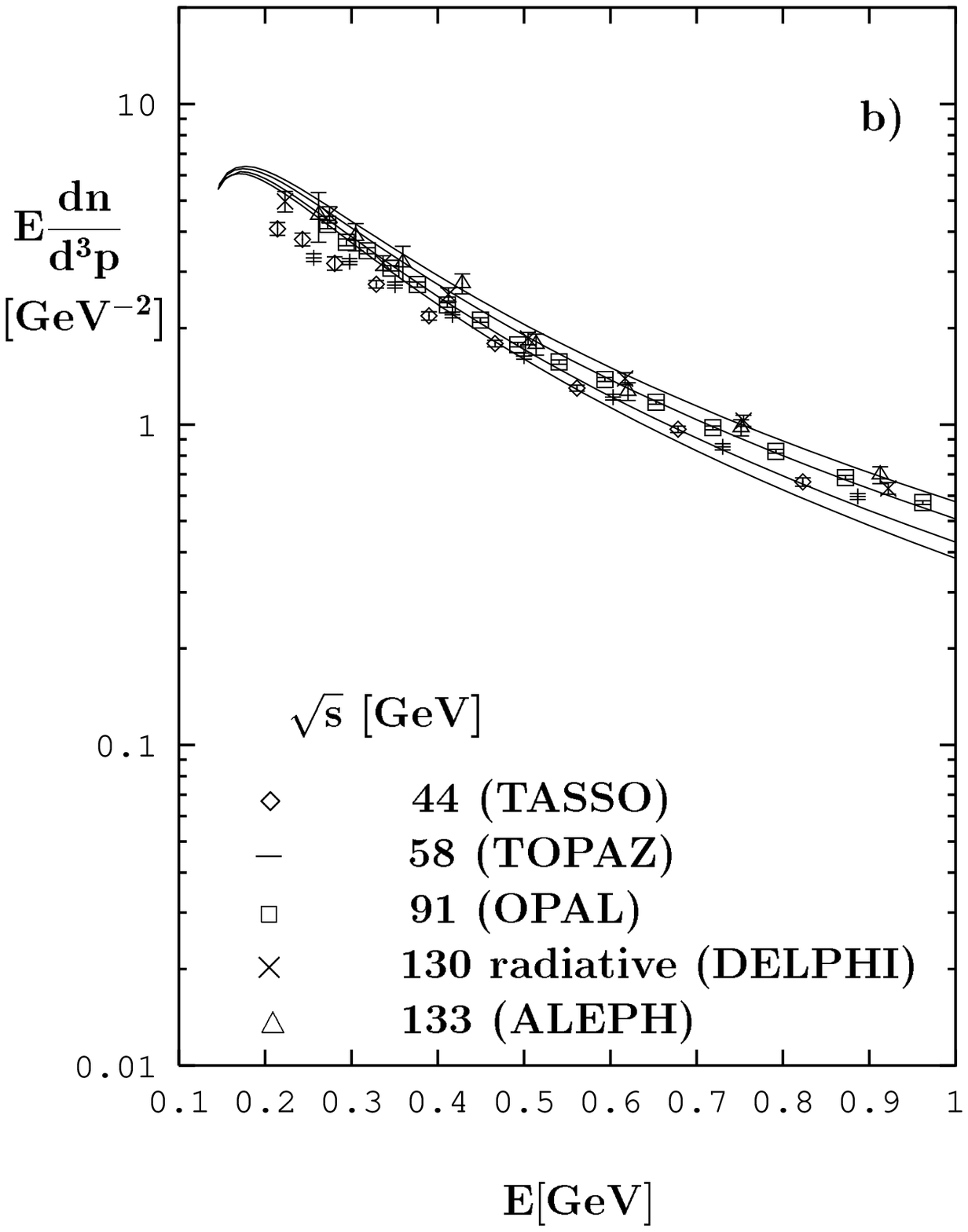,width=.6\linewidth,bbllx=5.5cm,bblly=10.5cm,bburx=13.5cm,bbury=26.5cm}}
          \end{center}
      \end{minipage}
\caption{{\bf a)} Invariant density $E dn/d^3p$ of charged particles 
in $e^+e^-$ annihilation 
as a function of the particle energy $E = \protect\sqrt{p^2 + Q_0^2}$ 
at $Q_0$ = 270 MeV at 
various $cms$ energies with $D_g^g$ computed using eq.~\protect\eref{phrel} 
($K_h$ = 0.45); 
{\bf b)} the same as in {\bf a)}, but at $Q_0$ = 138 MeV 
with $D_g^g$ computed from the Limiting Spectrum 
using eq.~\protect\eref{phrel1} ($K_h$ = 1.125).}
\end{figure}

Fig. 2a  shows 
the charged particle  invariant density, $E dn/d^3p$,  
as a function of the particle energy $E$ at different $cms$ energies 
ranging from 3 GeV up to LEP-1.5 $cms$ 
energy (133 GeV)\cite{data}
in comparison with the MLLA predictions for running $\alpha_s$ in the
approximate form\cite{klo1}. 
In this figure,  eq.~(\ref{phrel}) is used to relate parton
and hadron spectra at the effective mass 
$Q_0$ = 270 MeV which was found
\cite{lo} to provide a  good description of the moments of the
energy spectra over a large $cms$ energy interval. In Fig. 2b the same data are
compared with the MLLA limiting spectrum ($\lambda$ = 0), but 
 using eq.~\eref{phrel1}  with the mass value $Q_0=138$ MeV. 
Whereas this mass value is plausible for small energies $E$, the fit does
not work well for the full region of energies $E$ or the lower $cms$ energies. 
Note that, while these fits behave
differently very close to the boundary $E \simeq Q_0$, they clearly demonstrate 
the scaling behaviour expected from colour coherence. 
Furthermore they reproduce the
considerable energy dependence of the initial slope which can be 
 related to the
running of $\alpha_s$ in the MLLA damping factor.

It is remarkable that the data from all  $cms$ energies tend to converge in the
soft limit; we find $I_0 \simeq$ 6-8 GeV$^{-2}$ (using $Q_0$ = 270 MeV) and 
$I_0 \simeq$ 4-6 GeV$^{-2}$ (using $Q_0$ = 138 MeV). 
Inspecting the soft limit more closely, the 
LEP data seem to tend to a limiting value larger by about 20\% as compared to
the lower energy data. This may be well due to 
the overall  systematic effect in the relative normalization 
of the different experiments. 
Alternatively, there could be a contribution from incoherent sources,
for example, from weak kaon and heavy quark decays.

The good 
scaling behaviour of the soft limit $I_0$ is
also found \cite{klo2} in case of identified particles $\pi$, $K$, and $p$.
Remarkably, the scaling continues down to the lowest energy where data are
available, at $\sqrt{s}$=1.6 GeV \cite{adone}.

\section{Further tests of the parton hadron duality picture} 

It is suggestive to relate the observed limiting behaviour of particle spectra
for vanishing momenta to the expected behaviour of soft gluon emission off the
primary partons. 
The particle rate $I_0$ in this limit (see eq.~\eref{limit}) 
cannot be directly predicted, as
it depends on the normalization factor and the cut-off parameter $Q_0$.
However, the remarkable scaling behaviour of  $I_0$ 
in $e^+e^-$ annihilation provides one with a  standard scale for 
 the  comparison with other processes. 
 It would be very interesting to study whether the soft
 particle production is universal, i.e., a purely hadronic quantity  
 or whether the intensity $I_0$ indeed depends  
on the  colour topology  of the primary active partons in the collisions
process. This could provide one with 
a  direct support of the dual description of soft particle
production in terms of the QCD bremsstrahlung. 

In what follows we consider some possibilities to test this hypothesis further
and discuss what one may expect from the comparison of the $e^+e^-$ 
reaction with other collision processes. 

\subsection{gg Final State}
A direct test of the relevance of QCD coherence for the soft production limit is
the comparison of the $q\bar q$ with the $gg$ colour singlet final state. In
the $gg$ final state the same line of argument applies as above for the $q \bar
q$ final state, but the colour charge factor is increased by $N_C/C_F$ = 9/4 in
 eq.~\eref{limit} implying roughly a doubling of the soft 
 radiation intensity $I_0$.
An approximate
realization of the colour octet antenna is possible in $e^+e^- \to q \bar q g$
with the gluon recoiling against a quasi--collinear $q \bar q$
pair\cite{gary}. 

\subsection{3-Jet Events in $e^+e^-$ Annihilation} 
To be more quantitative, we consider the soft radiation in 3-jet $q\bar q g$
events of arbitrary jet orientation into a cone perpendicular to the production
plane and compare it to the radiation into the same cone in a 2-jet $q\bar q$
event again perpendicular to the primary $q\bar q$ directions. The analysis of
these configurations avoids the integration over the $k_T \ge Q_0$ boundaries
along the jets. 
We restrict ourselves to calculations of the soft gluon bremsstrahlung of
order $\alpha_s$. From the experience above with the two-jet events this
contribution dominates in the soft limit whereas higher order contributions
take over with increasing particle energy. 

Let us consider first the soft radiation into arbitrary direction $\vec{n}$
from a $q\bar q$ antenna pointing in directions $\vec{n}_i$ and
$\vec{n}_j$\cite{adkt}:
\be
dN_{q\bar q} = \frac{dp}{p} d\Omega_{\small\vec{n}} 
\frac{\alpha_s}{(2 \pi)^2} W^{q\bar q}(\vec{n}) \quad , \quad 
W^{q\bar q}(\vec{n}) = 2 C_F (\widehat{ij}) 
\label{wqq} 
\ee
with 
$(\widehat{ij}) = a_{ij}/( a_i a_j)$, $a_{ij} = (1 - \vec{n}_i
\vec{n}_j)$ and $a_i = (1 - \vec{n} 
\vec{n}_i)$. 
Such an antenna is realized, for example, in a $q \bar q \gamma$ event, and can
be obtained by the appropriate Lorentz boost from the $q\bar q$ rest frame.
The soft gluon radiation in a $q \bar q g$ event is given as in \eref{wqq} but
with the angular factor
\be
W^{q\bar q g}(\vec{n}) = N_C [ (\widehat{1+}) + (\widehat{1-}) 
- \frac{1}{N_C^2}  (\widehat{+-}) ] 
\ee
where $(+,-,1)$ refer to $(q,\bar q,g)$. 

For the radiation perpendicular to
the primary partons $(\widehat{ij}) = a_{ij} = 1 - \cos \Theta_{ij}$, with
 the relative angles $\Theta_{ij}$ between the primary partons $i$ and $j$. 
For 2-jet events of arbitrary orientation one obtains in this case 
\be
W^{q\bar q }_{\perp}(\Theta_{+-}) = 2 C_F ( 1 - \cos \Theta_{+-}) \;  .
\ee
Correspondingly the ratio $R_{\perp}$ of the soft particle yield in 3-jet events to
that of 2-jet events in their own rest frame 
($W_{\perp}^{q\bar q}(\pi) = 4 C_F$) is given by
\be
R_{\perp} \equiv \frac{dN_{\perp}^{q\bar q g}}{dN_{\perp}^{q\bar q}} =
\frac{N_C}{4 C_F} [ 2 - \cos \Theta_{1+} - \cos \Theta_{1-} - \frac{1}{N_C^2}
(1 - \cos \Theta_{+-} ) ]. 
\label{rperp}
\ee
This yields, in
particular, the limiting cases 
$R_{\perp}$ = 1 for soft or collinear primary gluon
emission ($\Theta_{1+}=1-\Theta_{1-}$, $\Theta_{+-}=\pi$)
and the effective $gg$ limit $R_\perp =N_C/N_F$
for the parallel $q \bar q$
($\Theta_{+-}$ = 0) configuration, as expected.
Eq. (\ref{rperp}) also predicts $R_\perp$ for all intermediate cases, in
particular, for Mercedes-type events 
one finds $R_\perp$=1.59. In this case 
no jet identification is needed for the above measurements. We also note
that the large angle gluon radiation is independent
of the mass of the quark (for
$\Theta \gg m_Q/ E_{jet}$). 

\subsection{Hard Processes with Primary Hadrons and Photons}

We consider here processes  with dominantly
2-jet final states. These include semihard processes which are initiated by 
partonic 2-body scatterings. In case of quark exchange 
(in particular, deep inelastic scattering processes $\gamma_Vp$,
$\gamma_V\gamma$ at large $Q^2$ in the Breit frame \cite{h1,zeus}) 
the two outgoing jets 
originate from the colour triplet charges and $I_0$ should be as in $e^+ e^-$
annihilation; in the case of gluon exchange $I_0$ should be 
about twice as large ($N_C/C_F$) as expected for  a $gg$ jet system.
In 2-jet events the limiting density $I_0$ may be most conveniently
determined as a limit of $dn/dyd^2p_T$ at $y\approx 0$ for $p_T\to
0$.
In general the deep inelastic processes may have contributions from 
both quark exchange and gluon exchange which are expected to dominate in
different frames.
Final states with several well separated jets can be
treated in analogy to the $e^+ e^- \to $ 3-jet events discussed before.

The frame dependence can be studied conveniently in these exchange
processes by considering $I_0(y)= \frac{dn}{dy d^2p_T} 
\bigr|_{p_T \to 0}$
as a function of rapidity $y$, measured, say, in the $cms$ frame. 
The Breit frame is reached by a Lorentz transformation in the direction of the
incoming photon, and corresponds to a different rapidity, $y_{Breit}$, in the
$cms$. One would then expect
for the processes with the virtual photons at large $Q^2$ 
in general a ``quark plateau" of
$I_0(y)$ in the current region of length $\Delta y \approx \log (Q/m)$
near $y_{Breit}$ and a transition to a 
``gluon plateau" in the complementary region of length 
$\Delta y \approx \log (W^2/Qm)$ where $m$ is a typical particle mass. The
existence of a plateau corresponds to the energy independence of $I_0$ as seen 
in $e^+e^-$ annihilation. So, if gluon exchange occurs with sufficient hardness
in the process, $I_0$ should develop a step like behaviour but never become
larger than twice the $I_0$ value in $e^+e^-$ annihilation. The height of the
``plateau'' is a direct indicator of the underlying exchange process (quark or
gluon like).

\subsection{Soft Collisions (Minimum Bias Events)}
These processes (with initial hadrons or real photons) are not so well
understood theoretically
as the hard ones but it might be plausible to extrapolate the gluon
exchange process towards small $p_T$\cite{ln}.
Experimental data at ISR energies, however, do not follow the expectation
of a doubling of $I_0$ in comparison to $e^+ e^-$ annihilation, rather the
intensities are found to be similar\cite{ppscal}. 
On the other hand, as shown in Fig.~3, 
$I_0$ roughly doubles when going from $\sqrt{s} \sim$ 20 GeV 
at S$p\bar p$S/ISR \cite{na22,ISR} 
to $\sqrt{s}$ = 900~GeV  at the S$p\bar p$S Collider\cite{coll}.  
If additional incoherent sources like weak decays can be excluded, such behaviour
could indicate the growing importance of one-gluon exchange expected from the
perturbative picture. Then a saturation at $I_0^{hh}/I_0^{e^+e^-} \sim 2$
is expected and no additional increase for a semihard $p_T$ trigger.
A rise of $I_0^{hh}$ could also result from incoherent multiple
collisions of partons (e.g. \cite{pythia}).  

\begin{figure}
          \begin{center}
\mbox{\epsfig{file=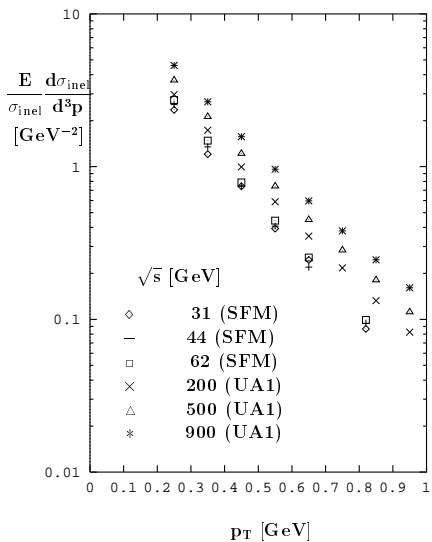,bbllx=4.5cm,bblly=10.cm,bburx=16.5cm,bbury=27.cm,height=7cm}}
       \end{center}
\caption{Invariant density $E dn/d^3p$ in hadronic collisions at central
rapidity as a function of  $p_T$ at different $cms$ energies.}
\end{figure}

\subsection{Production of $W$ pairs in $e^+e^-$ annihilation} 
If each of the W's decays into 2 jets the limiting soft particle yield 
will be twice the yield in $e^+e^- \to q \bar q$ , provided the interconnection
 phenomena\cite{interconn} are neglected. This tests directly the hypothesis of
independent emission.

\section{Conclusions}

The analytical perturbative approach to multiparticle production, based on the
MLLA and the LPHD,  
has proven to be successful in the description of various inclusive characteristics
of jets. It is of importance to investigate the limitations of this picture, in
particular in the soft region where non-perturbative effects are expected to occur.

It is remarkable that the intensity of the soft hadron production follows to a good
approximation a scaling law in a range of two orders of magnitude in the $cms$ energy
of $e^+ e^-$ annihilation (1.6-140 GeV). Such a scaling law is derived analytically for
the soft gluons in the jet and follows directly from the coherence of the soft gluon
radiation from all emitters. It appears  that the production of hadrons which is known
to proceed through many resonance channels nevertheless can be simulated in the average
through a parton cascade down to a small scale of a few 100 MeV as suggested by LPHD.
The scaling behaviour down to the very low $cms$ energy of 1.6 GeV can only
be explained within a parton model description if the cut-off scale $Q_0$
is well below 1 GeV.

It will be interesting to investigate the suggested scaling property~(1) 
and its possible violation further in the same experiment 
to avoid systematic
effects; this seems to be feasible at LEP and HERA for the simplest processes.
In particular, the effect of weak decays on scaling violations should be
clarified.
The sensitivity of the soft particle production to the effective colour 
charge of the primary emitters can be tested through the transverse 
production rates in multijet events.
The soft radiation in  $e^+ e^-$ annihilation can be used as a 
standard scale in the comparison of various processes. 
In this way the soft particle production can be a sensitive probe
of the underlying partonic process and the contributions 
from the incoherent sources.

\end{document}